\documentclass[journal=jacsat,manuscript=article]{achemso}

\usepackage{chemformula} 
\usepackage[T1]{fontenc} 
\usepackage{amsmath,amssymb,amsfonts,mathrsfs}
\usepackage{amsthm}
\usepackage{CJK}
\usepackage{bm}
\usepackage{lipsum}
\usepackage{babel}
\usepackage{graphicx}
\allowdisplaybreaks
\usepackage{makecell}
\usepackage{booktabs}
\usepackage{multirow}
\usepackage{gensymb}
\usepackage{rotating}
\usepackage{float}
\usepackage{braket}
\usepackage[breaklinks]{hyperref}
\usepackage{color}
\hypersetup{colorlinks=true, linkcolor=black, citecolor=black, filecolor=black, urlcolor=black}

\author{Quanchao Du}
\affiliation{%
 Ministry of Education Key Laboratory for Nonequilibrium Synthesis and Modulation of Condensed Matter, Shaanxi Province Key Laboratory of Advanced Functional Materials and Mesoscopic Physics, School of Physics, Xi'an Jiaotong University, Xi'an 710049, China 
}%

\author{Jinlian Lu}
\affiliation{%
 Department of Physics, Yancheng Institute of Technology, Yancheng, Jiangsu 224051, China
}%

\author{Xueqing Wan}
\affiliation{%
 Ministry of Education Key Laboratory for Nonequilibrium Synthesis and Modulation of Condensed Matter, Shaanxi Province Key Laboratory of Advanced Functional Materials and Mesoscopic Physics, School of Physics, Xi'an Jiaotong University, Xi'an 710049, China 
}%

\author{Zhenlong Zhang}
\affiliation{%
 Ministry of Education Key Laboratory for Nonequilibrium Synthesis and Modulation of Condensed Matter, Shaanxi Province Key Laboratory of Advanced Functional Materials and Mesoscopic Physics, School of Physics, Xi'an Jiaotong University, Xi'an 710049, China 
}%
\email{zhangzl@stu.xjtu.edu.cn}

\author{Zhijun Jiang}
\email{zjjiang@xjtu.edu.cn}
\affiliation{%
 Ministry of Education Key Laboratory for Nonequilibrium Synthesis and Modulation of Condensed Matter, Shaanxi Province Key Laboratory of Advanced Functional Materials and Mesoscopic Physics, School of Physics, Xi'an Jiaotong University, Xi'an 710049, China 
}%

\title{Ferroelectric Switchable Topological Magnon Hall Effect in Type-I Multiferroics}

\begin{document}
	\newpage
		
	\begin{center}
		\textbf{Abstract}
	\end{center}
	\baselineskip 22 pt
Electric control of magnetism at room temperature is crucial for developing next-generation, low-power spintronic devices. However, the intrinsic incompatibility between ferroelectricity and magnetism in crystal symmetry, along with the absence of strong magnetoelectric coupling mechanisms, continues to pose major challenges. In this work, we propose a general theoretical framework for magnon manipulation based on ferroelectric polarization switching in two-dimensional multiferroics. Taking monolayer multiferroics $\mbox{Ti}_{2}\mbox{F}_{3}$ as an example, our calculations demonstrate that ferroelectric switching can significantly modulate spin exchanges, thereby enabling nonvolatile and reversible electric control of the magnons. More importantly, the ferroelectric polarization reversal leads to a sign change in the Berry curvature, ensuring effective control over the valley Hall and nonlinear Hall response of magnons. This study provides a new way for realizing low-power and electrically controllable magnonic devices.
    ~~

	\newpage

	
\textit{Introduction.} 
Electric control of magnetism offers a most promising pathway for next-generation and high-performance spintronic devices, such as nonvolatile memory and low-power sensors\cite{ramesh2007multiferroics,fiebig2016evolution,schmid1994multi}. One promising approach is through the magnetoelectric (ME) coupling effect in multiferroics, where ferroelectric and magnetic orders coexist within a single phase\cite{spaldin2010multiferroics,xu2020electric,van2004origin,kimura2003magnetic,song2022evidence,ponet2022topologically}. Crucially, realizing the room-temperature multiferroics with strong ME coupling remains the significant challenges\,\cite{10.1093/nsr/nwz023}. In the so-called type-II multiferroics (e.g., $\mbox{TbMnO}_{3}$), although the intrinsic ME coupling is strong, the electric polarization induced by non-centrosymmetric spin order is typically quite small\cite{kimura2003magnetic,malashevich2008first,xiang2008spin}. Type-I multiferroics generally exhibit strong ferroelectricity and magnetism, but since these two orders originate from different mechanisms, the resulting ME coupling is usually weak\cite{spaldin2010multiferroics,wang2003epitaxial, PhysRevLett.122.117601}. Therefore, from both fundamental and application perspectives, discovering new mechanisms that enable efficient electrical control of magnetism is of great scientific importance and practical relevance. 

In addition to the spin orders, electric-field manipulation of spin excited states has recently emerged as a promising way. Particularly in magnetic insulators, magnons act as their charge-neutral quasiparticles of collective spin excitations, carrying spin angular momentum while effectively avoiding the Joule heating typically associated with charge transport\cite{chumak2015magnon,lenk2011building,onose_science_2010_329}. Therefore, achieving electric-field control of magnons holds significant potential for magnetic insulators. Recent studies have shown that an applied electric field can modulate the spin exchanges via the strong
spin-layer coupling\,\cite{,shen2024electrical, ni2025nonvolatile}. This mechanism enables nonvolatile control of magnons, including their topological and transport properties\cite{ni2025nonvolatile,wan2025topological,zhang2024predictable,tian2025spin}.
However, from the perspective of control efficiency, leveraging ferroelectric switching in multiferroics to manipulate the magnons represents a more effective approach, yet it remains largely unexplored.

\begin{figure}
    	\centering
    	\includegraphics[scale=0.65]{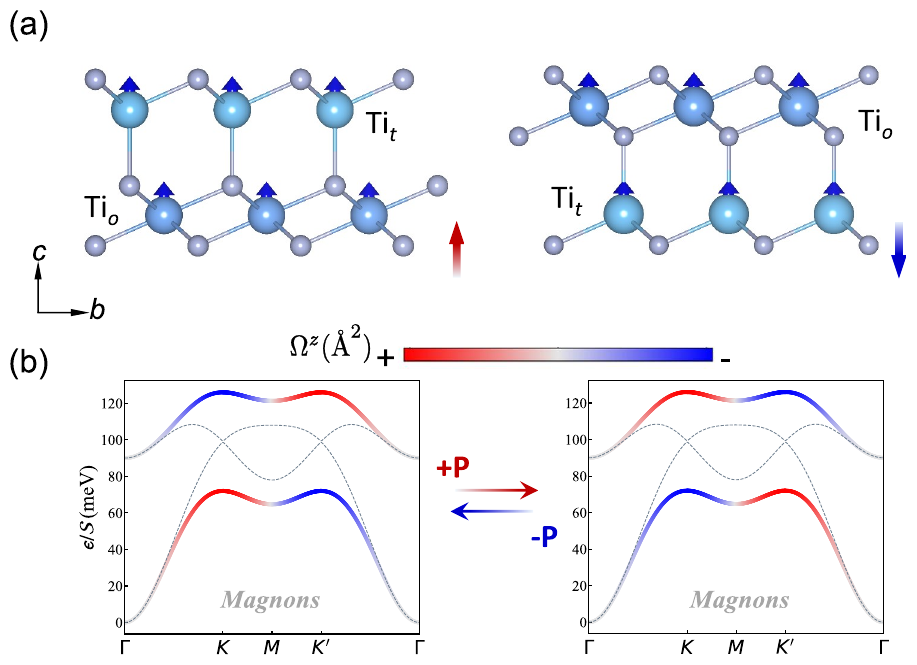}
    	\caption{(a) The side view of the monolayer $\mathrm{Ti_2F_3}$ showing different out-of-plane polarization $\mathrm{P}$ configuration, labeled as FE-up and FE-dn, respectively. (b) Magnon bands of  $\mathrm{Ti_2F_3}$ weighted by the  Berry curvature $\Omega^z$ for FE-up (left) and FE-dn (right) states. Switching the polarization from $+ \mathrm{P}$ to $- \mathrm{P}$ reverses the sign of $\Omega^z$. 
    \label{fig1}}
\end{figure}

In this work, we demonstrate that the ferroelectric-switching magnon Hall effect mechanism in Type-I multiferroics $\mathrm{Ti_2F_3}$\,\cite{zhang2025electric,sun2025anomalous}. Our results show that, the ferroelectricity in $\mathrm{Ti_2F_3}$ originate from the lattice distortion, which leads to the weak intrinsic ME coupling. However, ferroelectric polarization breaks the sublattice symmetry of the honeycomb ferromagnetic lattice, which ensure that ferroelectric switching is accompanied by a reversal of their magnon valley index. Consequently, ferroelectricity opens Dirac points for magnons at the valley points and induces nonzero Berry curvature. Since the Berry curvature determines the magnon valley Hall response, ferroelectric switching also reverses the corresponding valley transport. Importantly, although the weak spin–orbit coupling $\mathrm{Ti_2F_3}$ leads to a negligible linear thermal Hall response, the nonlinear Hall effect arising from inversion symmetry breaking is highly robust and can likewise be efficiently controlled through ferroelectric switching.

\begin{figure}
    	\centering
    	\includegraphics[scale=0.65]{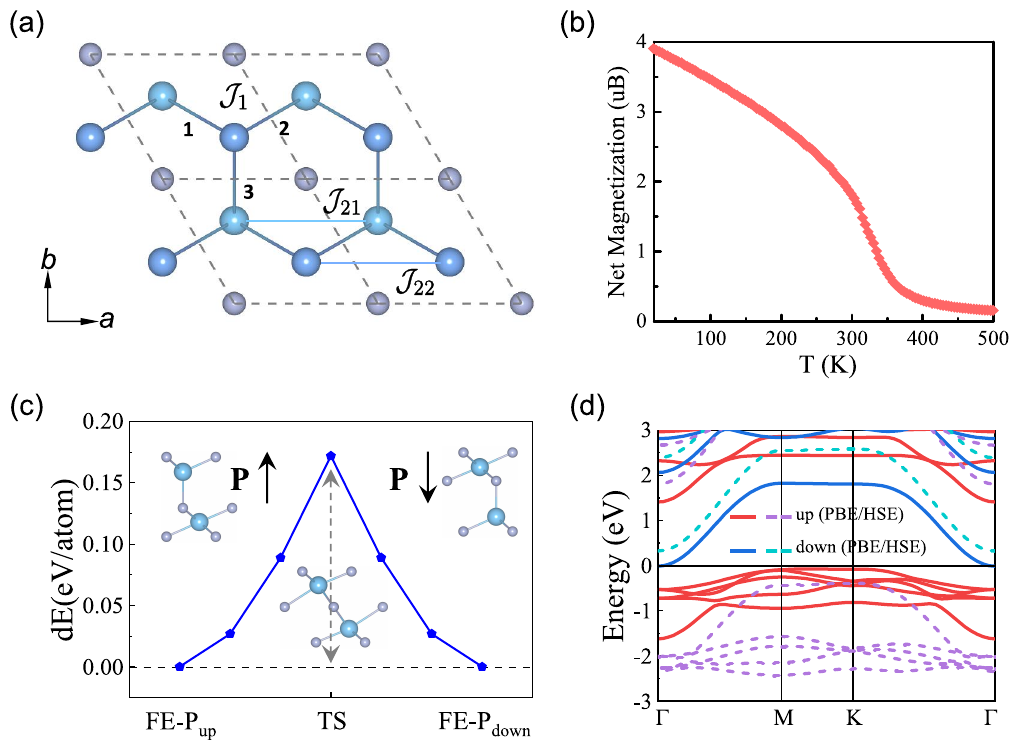}
    	\caption{(a) Top view of the monolayer $\mathrm{Ti_2F_3}$ structure with labeling the spin exchanges. Here, light and dark blue ions refer to the $\mathrm{Ti}_{t}$ and
        $\mathrm{Ti}_{o}$, respectively. (b) Temperature dependence of the net magnetization $\mathrm{Ti_2F_3}$ obtained from MC simulations. (c) The FE switching barrier of monolayer $\mbox{Ti}_{2}\mbox{F}_{3}$. (d) The band structure of the monolayer $\mbox{Ti}_{2}\mbox{F}_{3}$.
    \label{fig2}}
    \end{figure}

\textit{Magnetic properties of $\mathrm{Ti_2F_3}$.} Monolayer $\mathrm{Ti_2F_3}$ adopt the $P3m1$ polar space group, with its structure illustrated in Figure\,\ref{fig1}(a). The ferroelectricity in monolayer $\mathrm{Ti_2F_3}$ originates from the charge transfer between the Ti magnetic ions and F anions. This charge redistribution drives the system into an insulating state and simultaneously induces lattice distortions in the Ti-F sub-layer, as shown in Figure\,\ref{fig2}(c)-(d). Consequently, in ferroelectric-up configuration (FE-up), the upper Ti-F sub-layer adopts a $\mathrm{TiF_{4}}$ tetrahedral geometry ($\mathrm{Ti}_{t}$), whereas the down Ti-F sub-layer exhibits a $\mathrm{TiF_6}$ octahedral coordination environment ($\mathrm{Ti}_{o}$)\,\cite{zhang2025electric,sun2025anomalous}. In the ferroelectric-down (FE-dn) configuration, this arrangement is reversed. The stability of the ferroelectric monolayer $\mathrm{Ti_2F_3}$ is confirmed by the first principles phonons calculations, as shown in Figure\,S1\,\cite{Supplemental_Materials}. The energy barrier between the FE-up and Fe-dn state is also calculated by the Nudged-Elastic-Band (NEB) theory\cite{sheppard2012generalized} (see Figure\,\ref{fig2}(c)). Based on their relative positions, $\mathrm{Ti}$ magnetic ions in the unit cell can be distinguished into two sublattices, $\mathrm{Ti}_{t}$ and $\mathrm{Ti}_{o}$, which are convertible upon ferroelectric switching. 

Evidently, the top view of the Ti magnetic ions in monolayer $\mathrm{Ti_2F_3}$ forms a ferromagnetic honeycomb lattice, as shown in Figure\,\ref{fig2}(a). In this context, nearest-neighbor (NN) intralayer spin exchange between $\mathrm{Ti}$ ions corresponds to NN spin exchange of the honeycomb lattice, while NN spin exchange between $\mathrm{Ti}$ magnetic ions corresponds to second-nearest-neighbor (2NN) spin exchange. DFT calculations reveal that all NN spin exchanges between the Ti magnetic ions exhibit ferromagnetic character. Since the weak spin-orbit coupling, the spin Hamiltonian for the $\mathrm{Ti_2F_3}$ monolayer can be well described by the Heisenberg model, given by 
\begin{equation}\label{eq1}
\begin{split}{\cal \hat{H}}=  {\cal J}_{1}\sum_{\langle i,j\rangle}{\cal S}_{i}\cdot{\cal S}_{j}+{\cal J}_{21}\sum_{\langle i,j\rangle}{\cal S}_{i}\cdot{\cal S}_{j}+{\cal J}_{22}\sum_{\langle i,j\rangle}{\cal S}_{i}\cdot{\cal S}_{j} ,
\end{split}
\end{equation}
where ${\cal J}_{1}$ is the NN interlayer spin exchange between $\mathrm{Ti}_{o}$ and $\mathrm{Ti}_{t}$, ${\cal J}_{21(22)}$ denotes NN intralayer spin exchange between $\mathrm{Ti}_{o}$-$\mathrm{Ti}_{o}$ ($\mathrm{Ti}_{t}$-$\mathrm{Ti}_{t}$) ions. Based on the LKAG method\,\cite{liechtenstein1987local, grotheer2001fast, he2021tb2j}, the ${\cal J}_{1}$, ${\cal J}_{21}$, and ${\cal J}_{22}$ is obtained, with values of that -$\mbox{30}$, -$\mbox{10}$ and -$\mbox{5}$ $\mbox{meV}$, respectively. The Monte Carlo (MC) simulations\,\cite{Supplemental_Materials}, as presented in Figure\,\ref{fig2}(b), indicate that ferromagnetic transition temperature ($\mathrm{T_c}$) of approximately 340\,$\mbox{K}$, which is above room temperature.

Within the ferromagnetic order, the linear spin wave (LSW) model in Eq.(\ref{eq1}) can be solved by employing the Holstein-Primakoff (HP)\citep{holstein1940field}, ${\cal S}_{i,\alpha}^{z}=S-\hat{a}_{i,\alpha}^{\dagger}\hat{a}_{i,\alpha},{\cal S}_{i}^{+}\approx\sqrt{2S}\hat{a}_{i,\alpha}$, and ${\cal S}_{i}^{-}\approx\sqrt{2S}\hat{a}^{\dagger}_{i,\alpha}$, with $\hat{a}_{i,\alpha}^{\dagger}\,(\hat{a}_{i,\alpha}^{\dagger})$ creating a magnon at sublattice in the honeycomb unit cell. After the Fourier transformation, the Hamiltonian can be expressed in momentum space using the basis $\psi_{k}^{\dagger}\equiv(\hat{a}_{k}^{\dagger},\hat{b}_{k}^{\dagger})$ as ${\cal {\hat{H}}}=\sum_{k}\psi_{k}^{\dagger}{\cal \hat{H}}_{k}\psi_{k}$. Neglecting the zero-point energy, the ${\cal \hat{H}}_{k}$ reads as,
\begin{equation}\label{eq2}
\frac{{\cal \hat{H}}_{k}}{S} =-3{\cal J}_{1}+\left(\begin{matrix}{\cal J}_{21}f_{k} & {\cal J}_{1}\gamma_{k}\\
{\cal J}_{1}\gamma_{k}^{\dagger} & {\cal J}_{22}f_{k}
\end{matrix}\right),
\end{equation}
where $\gamma_{k}=\sum_{\delta}e^{-i\textbf{\textit{k}}\cdot\boldsymbol{\delta_{i}}}$ and $f_{k}=\sum_{i\in odd}2\mbox{cos}(\textbf{\textit{k}}\cdot\boldsymbol{\mu}_{i})-6$ with $\boldsymbol{\delta}_{i}$ and $\boldsymbol{\mu}_{i}$ being NN and 2NN linking vectors of honeycomb lattice as shown in Figure\,\ref{fig2}({a}).

\begin{figure}
		\centering
		\includegraphics[scale=0.6]{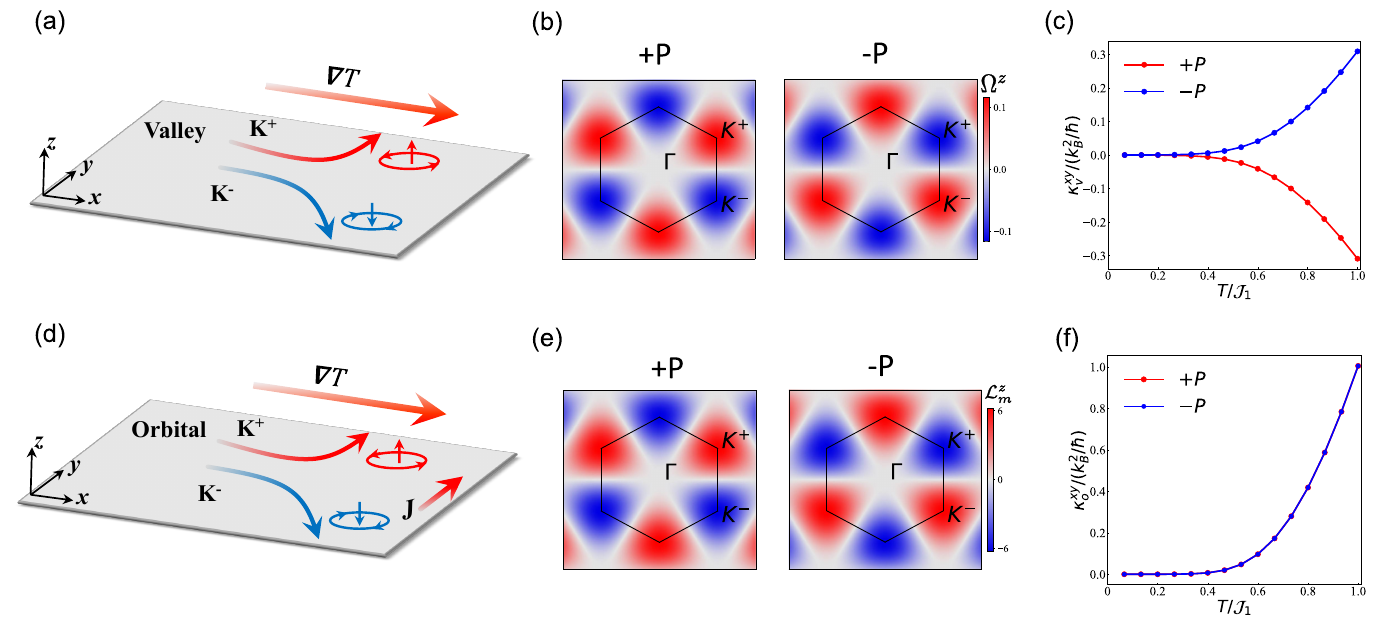}
		\caption{(a) The schematic of the magnon valley Hall effect. (b) The Berry curvature of the major magnon band with FE-up (left) and FE-dn (right) state. (c) The temperature dependence of the magnon valley Hall conductivity ${\kappa}^{xy}_{v}$. (d) The schematic of the magnon orbital Hall effect. (e) The orbital magnetic moment of the major magnon band with FE-up (left) and FE-dn (right) state. (f) The temperature dependence of the magnon orbital Hall conductivity ${\kappa}^{xy}_{o}$. 
			\label{fig3}}
\end{figure}

We further parameterize Eq.\,(\ref{eq2}) as $h_{0}=\frac{1}{2}({\cal J}_{21}+{\cal J}_{22})f_{k}$, $h_{x}={\cal J}_{1}\mbox{Re}\gamma_{k}$, $h_{y}={\cal J}_{1}\mbox{Im}\gamma_{k}$ and $h_{z}=\frac{1}{2}({\cal J}_{21}-{\cal J}_{22})f_{k}$. Eq.\,(\ref{eq2}) can thus be expressed in terms of the Pauli matrices $\boldsymbol{\sigma}=\left(\sigma_{x},\sigma_{y},\sigma_{z}\right)$ as
\begin{equation}\label{eq3}
\frac{{\cal \hat{H}}_{k}}{S}=h_{0}I+\boldsymbol{h}(\boldsymbol{k})\cdot\boldsymbol{\sigma},
\end{equation}
where $\boldsymbol{h}(\boldsymbol{k})=\left(h_{x},h_{y},h_{z}\right)=|h|\left(\mbox{sin}\theta\mbox{cos}\phi,\mbox{sin}\theta\mbox{sin}\phi,\mbox{cos}\theta\right)$. As a result, the corresponding eigenvalues and eigenvectors are given as 
\begin{equation}\label{eq4}
\frac{{\epsilon}_{\pm}}{S}=h_{0}\pm|\boldsymbol{h}(\boldsymbol{k})|,\quad\Psi_{\pm}=\begin{pmatrix}\sqrt{1\pm\mbox{cos}\theta}\\
\pm e^{-i\phi}\sqrt{1\mp\mbox{cos}\theta}
\end{pmatrix}.
\end{equation}
The presence of $h_{z}$ explicitly breaks the sublattice symmetry of the honeycomb lattice, opening a gap in the magnon bands at the $\mbox{K}$-points. While time-reversal symmetry ($\cal{T}$) remains preserved\cite{mook_PRX_2021_11, wan2025topological}, the broken inversion symmetry ($\cal{P}$) leads to a nonzero Berry curvature. In 2D case, the Berry curvature only has $z$ component\cite{ni2025nonvolatile, xiao2010berry}, which can be expressed in terms of the $\boldsymbol{h}(\boldsymbol{k})$ vector 
\begin{equation}\label{eq5}
\boldsymbol{\Omega}_{\pm}^{z}(\boldsymbol{k}) =-i\langle\boldsymbol{\nabla}\Psi_{\pm}|\times|\boldsymbol{\nabla}\Psi_{\pm}\rangle =\mp\frac{h_{z}}{2}\left(\boldsymbol{\nabla}h_{x}\times\boldsymbol{\nabla}h_{y}\right).
\end{equation}
Therefore, the nonzero Berry curvature of magnons in monolayer $\mathrm{Ti_2F_3}$ can be emerged in the ferroelectric states, in which nonzero of ${\cal J}_{21}$\,$-$\,${\cal J}_{22}$ breaks the ${\cal P}$ symmetry, leading to that $\boldsymbol{\Omega}(\boldsymbol{k})=-\boldsymbol{\Omega}(-\boldsymbol{k})$. Notably, the sign of ${\cal J}_{21}$\,$-$\,${\cal J}_{22}$ can be reversed through ferroelectric switching, enabling electrical control of the magnonic Berry curvature, as shown in the Figure\,\ref{fig1}(b) and Figure\,\ref{fig3}(b).

In addition to the Berry curvature, the emergence of $h_{z}$ can induce orbital moments of magnons\cite{neumann2020orbital, fishman2022orbital, go2024magnon, ni2025nonvolatile}.
According to the previous work\cite{chang1996berry, xiao2010berry}, the orbital quantity of Bloch magnons can be expressed as
\begin{equation}\label{eq6}
\boldsymbol{{\cal L}}_{m,\pm}^{z}(\boldsymbol{k})=-\frac{i}{2\hbar}\langle\boldsymbol{\nabla}\Psi_{\pm}|\times(\hat{{\cal H}}_{k}-\epsilon_{k})|\boldsymbol{\nabla}\Psi_{\pm}\rangle.
\end{equation}
As shown in Figure\,\ref{fig3}(e), orbital moments (${\cal L}_{m}^{z}$) in the ferroelectric state mainly locate at the valley points, and their sign can be reversed through ferroelectric switching, similar to the behavior of the Berry curvature. Notably, the orbital moments exhibit the same chirality for both the majority and minority magnon bands. Therefore, in monolayer $\mathrm{Ti_2F_3}$, the ferroelectricity breaks the ${\cal P}$ symmetry of the honeycomb lattice, inducing finite Berry curvature and orbital magnetic moments in magnons, whose signs can be reversibly controlled through ferroelectric switching.

\textit{Magnon valley and orbital response.} As shown in Figure\,\ref{fig3}, the Berry curvature of magnons with ferroelectric state is primarily concentrated at the ${\mbox{K}}$ points, satisfying $\boldsymbol{\Omega}(\boldsymbol{k})=-\boldsymbol{\Omega}(-\boldsymbol{k})$. In this case, the linear thermal Hall response under the temperature gradient remains zero due to the opposite contributions of magnon Berry curvature at $\mbox{K}$ and $\mbox{K}^{\prime}$-points. However, the opposite orbit pseudo-magnetic field at $\mbox{K}$ and $\mbox{K}^{\prime}$-points gives rise to the magnon valley current, similar to the gapped graphene\cite{xiao2007valley, bhowal2021orbital}. To distinguish
the difference between $\mbox{K}$ and $\mbox{K}^{\prime}$-points \cite{zhai2020topological},
we define the magnon valley conductivity ${\cal \kappa}^{v}_{xy}$ as 
\begin{equation}\label{eq7}
{\cal \kappa}_{xy}^{v}=\frac{k_{B}^{2}T}{\hbar V}\sum_{n,k}c_{2}(\rho)\left[\Omega_{n,k}(\mbox{K})-\Omega_{n,k}(\mbox{K}^{\prime})\right],
\end{equation}
where $V$ is the volume, $\varepsilon_{abc}$ is the Levi-Civita
symbol with $abc=xyz$ and $c_{2}(\rho)=\int_{0}^{\rho}[\mbox{log}(1+\rho^{-1})]^{2}d\rho$ with $\rho$ being Boltzmann distributions for bosons. Since the Berry curvature can be reversed by the ferroelectric switching, the sign of the magnon valley Hall conductivity ${\cal \kappa}_{xy}^{v}$ can be controlled by the ferroelectric polarization, as illustrated in Figure\,\ref{fig3}(c).

Next, we investigate the magnon orbital response in the ferroelectric $\mathrm{Ti_{2}F_{3}}$. The emergence of magnon orbital moments gives rise to the anomalous orbital associated transport phenomena\cite{go2024magnon, bhowal2021orbital}, in which the orbital Berry curvature, $\Omega_{o}^{z}$ is defined as 
\begin{equation}\label{eq8}
\boldsymbol{\Omega}_{o}^{z}(\boldsymbol{k})=\boldsymbol{\Omega}^{z}(\boldsymbol{k})\boldsymbol{{\cal L}}_{m}^{z}(\boldsymbol{k}).
\end{equation}
Obviously, the orbital Berry curvature is the product of the Berry
curvature and the orbital moments of magnons\cite{go2024magnon, bhowal2021orbital}.
As both the Berry curvature and orbital moments are odd with respect
to ${\boldsymbol{k}}$, their product, orbital Berry curvature, is
even, $\boldsymbol{\Omega}_{o}^{z}(\boldsymbol{k})=\boldsymbol{\Omega}_{o}^{z}(-\boldsymbol{k})$. This feature enables the emergence of the magnon orbital Hall response, in contrast to the contribution from the Berry curvature alone. Analogous to the magnon Hall effect,
the magnon orbital Hall conductivity can be determined by integrating the product of $c_{2}(p)$ and $\boldsymbol{\Omega}^{z}_{o}$ over the Brillouin zone, expressed as follows 
\begin{equation}\label{eq9}
{\cal \kappa}_{xy}^{o}=\frac{k_{B}^{2}T}{\hbar V}\sum_{n,k}c_{2}(\rho)\Omega_{n,o}^{z}(k). 
\end{equation}
As shown in Figure\,\ref{fig3}(f), we calculated the temperature-dependent magnon orbital Hall conductivity. Obviously, unlike the magnon valley Hall response, the orbital Hall conductivity remains unchanged across different ferroelectric states. This difference mainly arises from the fact that, although both the Berry curvature and the orbital moment can be reversed by ferroelectric polarization, their product—the orbital Berry curvature—remains invariant.

\begin{figure}
		\centering
		\includegraphics[scale=0.75]{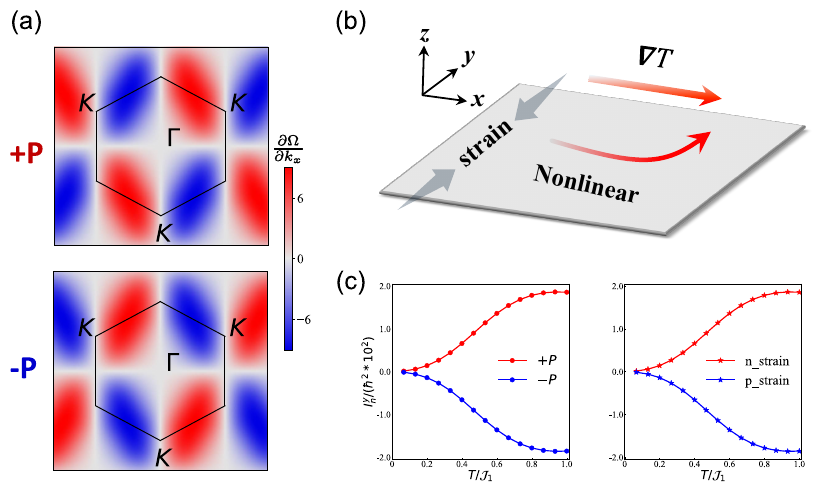}
		\caption{(a) The extended BCD of the major band with FE-up and FE-dn state.  (b) The schematic illustration of the magnon nonlinear Hall effect. (c) The calculated temperature dependence of the magnon nonlinear Hall conductivity ${I}^{y}_{n}$. The left and right panels illustrate the reversal behavior induced by ferroelectric polarization switching and strain modulation, respectively.
			\label{fig4}}
\end{figure}

\textit{Magnon nonlinear Hall response.} Additionally, the ${\cal P}$ breaking in magnons gives rise to the nonlinear Hall response. According to the previous works\,\cite{ni2025magnon, mukherjee_PRB_2023_107, kondo_prreserach_2022_4}, the magnon nonlinear Hall conductivity is given as  
\begin{equation}\label{nonlinear_1}
\begin{split}
    I_{n}^{a} \approx \frac{2\varepsilon_{abc}} { V\hbar^{2} T}  \sum_{n,k} c_{1}(\rho) \epsilon_{n,k}{\partial_{k_{b}} \left[ {\Omega^{c}_{n,k}{\epsilon}_{n,k}} \right]}, 
\end{split}
\end{equation}
where $\varepsilon_{abc}$ is Levi-Civita symbol with $abc=xyz$ and $c_{1}(\rho) = \int^{\rho}_{0} \mbox{log}(1+\rho^{-1}) d\rho $. In 2D case, $c$\,$=$\,$z$, indicating that the nonlinear transverse magnon current can emerge along $y$-axis when applying $\boldsymbol{\nabla}T$ along $x$-axis as shown in Figure\,\ref{fig4}(b). Obviously, $ I_{n}^{y}$ depends on the ${\partial_{k_{x}}}{\Omega_{n,k}}$, which is defined as the extended Berry curvature dipole (BCD) of magnons\,\cite{kondo_prreserach_2022_4, mukherjee_PRB_2023_107}. The extended BCD distribution of major band is shown Figure\,\ref{fig4}(a), which is even with respect to $k$ and can be reversed by the ferroelectric switching. Notably, since the ${\cal C}_{3z}$ symmetry can transform $k_{m}$ to $k_{\mu}$ and $k_{\nu}$ in $k$-space, allows
$\left[{\partial_{k_{x}}}{\Omega_{n,}}\right]_{k_{m}}$\,$+$\,$\left[{\partial_{k_{x}}}{\Omega_{n,}}\right]_{k_{\mu}}$\,$+$\,$\left[{\partial_{k_{x}}}{\Omega_{n,}}\right]_{k_{\nu}}$\,$=$\,$0$,
suggesting the integral of Eq.\,(\ref{nonlinear_1}) goes to zero. As shown in Figure\,\ref{fig4}(c) and Figure\,S3\,\cite{Supplemental_Materials}, the nonzero sum of velocities can be generated by uniaxial strain, which breaks the ${\cal C}_{3z}$ symmetry and distributes the invariance among the three NN spin exchange interactions. When a little uniaxial strain perpendicular to the applied temperature gradient is introduced into the monolayer $\mathrm{Ti_{2}F_{3}}$, the pronounced nonlinear transverse magnon current can be emergent. Interestingly, in addition to ferroelectric switching, the direction of the applied uniaxial strain is also coupled to the sign of the nonlinear Hall response. As shown in Figure\,S3\,\cite{Supplemental_Materials} and Figure\,\ref{fig4}(c), the nonlinear Hall response can be reversed by either the ferroelectric switching or the uniaxial strain.

\textit{Summary and discussion.} In contrast to the linear magnon Hall response, which typically requires breaking the effective time-reversal symmetry via the Dzyaloshinskii-Moriya interaction (DMI) or bond-dependent spin exchanges\,\cite{katsura_prl_2010_104, onose_science_2010_329, zhang_thermal_Physreport_2024_1070, owerre2016first}, the nonlinear Hall response in $\mathrm{Ti_{2}F_{3}}$ arise from the lattice distortions. Importantly, both the ferroelectric and strain switching can reverse the sign of the nonlinear Hall conductivity. Therefore, although monolayer $\mathrm{Ti_{2}F_{3}}$ exhibits weak spin-orbit coupling and a negligible linear Hall response, its robust nonlinear Hall effect, combined with tunability via external fields, endows it with significant potential for practical applications.

In summary, we reveal the mechanism by which ferroelectric switching modulates the topological magnon transport in monolayer multiferroics $\mathrm{Ti_{2}F_{3}}$. Our results show that the ferroelectric distortion breaks the inversion symmetry of the hexagonal ferromagnetic lattice, enabling the Berry curvature and orbital magnetic moment of magnons to reverse with the ferroelectric polarization. Consequently, both the valley Hall transport and the nonlinear Hall effect are strongly coupled to the ferroelectric polarization. These findings provide important theoretical guidance for realizing high-performance and reconfigurable next-generation spintronic devices.
    
	~~\\
	~~\\
	\textbf{ASSOCIATED CONTENT}\\
	\textbf{Supporting Information}\\
    In this Supporting Information, we provide: I. DFT calculations and Monte Carlo simulations on monolayer $\mathrm{Ti_2F_3}$; II. Topological geometry of magnons in monolayer $\mbox{Ti}_{2}\mbox{F}_{3}$.

    ~~~\\
	~~\\
	\textbf{Corresponding Authors}\\
    	$^*$E-mail: zhangzl@stu.xjtu.edu.cn (Z. Z.)\\
    $^*$E-mail: zjjiang@xjtu.edu.cn (Z. J.)\\
	\textbf{Notes}\\
	The authors declare no competing financial interest. 
	
	~~~\\
	~~~\\
	\textbf{ACKNOWLEDGMENTS}\\
The authors thank Prof.\,Jianli Wang and Prof.\,Jinyang Ni for helpful discussions. This work is supported by the National Natural Science Foundation of China (Grant No. 12374092), Natural Science Basic Research Program of Shaanxi (Program No. 2023-JC-YB-017), Shaanxi Fundamental Science Research Project for Mathematics and Physics (Grant No. 22JSQ013), “Young Talent Support Plan” of Xi'an Jiaotong University, and the Xiaomi Young Talents Program.\\

\bibliography{Main}

\end{document}